# Breaking Bad News in the Era of Artificial Intelligence and Algorithmic Medicine:

*An Exploration of Disclosure and its Ethical Justification using the Hedonic Calculus*

**Authors**: Benjamin Post[1,2,3], Cosmin Badea[2], Aldo Faisal[1,2,3,5], Stephen J. Brett[3,4]

**Affiliations:**
1 Department of Bioengineering, Imperial College London, London, UK
2 Department of Computing, Imperial College London, London, UK
3 UKRI Centre in AI for Healthcare, Imperial College London, London, UK;
4 Department of Surgery and Cancer, Imperial College London, UK
5 Institute of Artificial & Human Intelligence, University of Bayreuth, Germany

*Corresponding Author:* Benjamin Post b.post19@imperial.ac.uk
*Pre-print submitted by:* Cosmin Badea c.badea@imperial.ac.uk

**ORCID IDs:**
BP - 0000-0003-0337-0827
CB - 0000-0002-9808-2475
AF - 0000-0003-0813-7207
SJB- 0000-0003-4545-8413

**Statements and declarations:**

BP was supported by the UKRI CDT in AI for Healthcare http://ai4health.io (Grant No. P/S023283/1). AAF holds a UKRI Turing AI Fellowship (Grant No. EP/V025449/1). This research was also funded by the NIHR Imperial Biomedical Research Centre (BRC). The views expressed are those of the authors and not necessarily those of the NIHR or the Department of Health and Social Care

For the purpose of open access, the author has applied a 'Creative Commons Attribution (CC BY) licence to any Author Accepted Manuscript (AAM) version arising.

None of the authors have any conflicts of interest to declare

**Abstract** (150-250 words): 163

An appropriate ethical framework around the use of Artificial Intelligence (AI) in healthcare has become a key desirable with the increasingly widespread deployment of this technology. Advances in AI hold the promise of improving the precision of outcome prediction at the level of the individual. However, the addition of these technologies to patient-clinician interactions, as with any complex human interaction, has potential pitfalls. While physicians have always had to carefully consider the ethical background and implications of their actions, detailed deliberations around fast-moving technological progress may not have kept up. We use a common but key challenge in healthcare interactions, the disclosure of bad news (likely imminent death), to illustrate how the philosophical framework of the 'Felicific Calculus' developed in the 18$^{th}$ century by Jeremy Bentham, may have a timely quasi-quantitative application in the age of AI. We show how this ethical algorithm can be used to assess, across seven mutually exclusive and exhaustive domains, whether an AI-supported action can be morally justified.

| Felicific Variables | Definition |
|---|---|
| Intensity | *The intensity of the pleasure or pain* |
| Duration | *How long the pleasure or pain will last* |
| Certainty | *The probability that the pleasure or pain will occur* |
| Propinquity | *How soon the pleasure or pain will occur* |
| Fecundity | *How likely the sensation (pleasure or pain) is to lead to more of the same sensation* |
| Purity | *How likely the sensation (pleasure or pain) is to lead to the opposite sensation* |
| Extent | *The number of people affected by the pleasure or pain* |

**Table 1:** The Domains of the Felicific Calculus

# 1 Introduction

A great deal of effort is currently being expended on developing risk prediction models for individuals and patient groups using a variety of approaches ranging from genomics and metabonomics through to socioeconomic phenotyping[1]–[6]. In the domain of healthcare, the expansion in predictive modelling research is paired with rapidly emerging concerns about the ethical use of such methods, particularly artificial intelligence (AI)[7], [8]. These include concerns around data privacy, algorithmic fairness, bias, safety, informed consent, and transparency[7], [9]–[11], for which the medical profession may be unprepared to navigate[11]. Accordingly, international bodies have started taking action to address concerns around medical AI and automation. Last year, in their extensive report regarding AI in healthcare, the World Health Organisation explicitly stated that 'humans should remain in full control' of medical decisions[12]. Article 22 of the European Union's General Data Protection Regulation (GDPR) outlines the right of individuals not to be subject to decisions "based solely on automated processing"[13]. The importance of keeping humans involved in significant medical decisions needs to be carefully considered as statistical, machine learning and artificial intelligence models are increasingly aiding diagnostics, treatment decisions and outcome prediction[8].

The modelling of future life-threatening events and of death is one of the most common applications of predictive tools in healthcare, which well-encapsulates the ethical issues above, as well as the complexity of individualised prediction. Furthermore, these predictions may have significant implications for patients and the communication of this highly sensitive medical information has its own ethical challenges. To assess whether complex actions are justified, *medical ethics* is highly integrated into healthcare practice. *Utilitarianism*, which prioritises maximising benefit to the greatest number of people, is a well-established paradigm within medical ethics that provides a tool to comprehensively consider an ethical dilemma – the *Felicific Calculus*[14]. This is a useful approach for contemplating complex medical interventions with the potential to impact many people [15], [16].

We thus set out to explore the consequences of disclosing bad news in the era of AI from a Utilitarian point of view. We provide background information on the current use of AI in healthcare, the complexity of disclosing bad news and an outline of the Felicific calculus. We move on to using this ethical tool to systematically investigate whether this disclosure can be deemed good or bad and finish by discussing the relevance of our findings in a rapidly evolving domain.

## 2  Background

**AI Predictions in Healthcare**

The current situation is that artificial intelligence (AI) has the potential to benefit patients by using precise and individualised modelling[17]–[19], but new risks and complexity need to be carefully considered[12]. These AI-predictive models may be used to predict treatment outcomes, life-expectancy and progression of rehabilitation[20], [21]. Communicating these predictions will have an impact on patients, loves ones, carers, as well as healthcare workers. Complications can arise as most AI-based predictive models are trained on datasets comprised of numerous individuals' data, allowing for larger sample sizes, more training data and better population-level performance. However, when these models are applied at the *level of the individual* significant uncertainty may be carried forward[22], [23]. So, how should a patient or clinician interpret a disease or outcome prediction made by a mathematical model, derived from many other peoples' healthcare data?

A complex relationship exists between the quality of a model's prediction and the ability or desire of healthcare providers to act on this knowledge. If an algorithm can accurately identify an individual at risk of a serious and preventable or treatable condition, our duty to disclose this information arguably increases. However, if diagnostic predictions are made that have poor precision, identify asymptomatic diseases or those that cannot be cured or medically acted upon, the moral imperative may be less clear.

We offer here a way forward, taking a view from medicine and artificial intelligence, we speculate that the outputs of prediction modelling, in general, should be interpreted and presented as a *risk* or *probability* of an event occurring to an individual rather than a certainty[24]–[26]. Additionally, probabilistic methods and more empirical approaches such as calibration testing can provide a *confidence* (or uncertainty) around such predictions and arguably these should also be presented. Yet, every model is limited and biased by the data that were used to train it, with modelling assumptions and factors simply not captured or known at the time of development (e.g. due to persistently changing conditions). Furthermore, algorithms developed from specific patient cohorts may not translate well to populations in different parts of the world, with different demographics or baseline medical conditions[20], [23], [27]–[29]. These are complex concepts and not necessarily intuitive, even to experts in clinical and technical disciplines. The ethical implications of this knowledge may be uncertain and, hence, the appropriateness of disclosing this information may not be straightforward.

**Disclosing Bad News**

In the healthcare context, 'bad news' can be defined as information that creates a negative view of a person's health[30] or reduces their choices in life[31], [32]. Historically the protection of patients from potentially distressing news was regarded as reasonable and consistent with a physician's role[33]. This strongly paternalistic approach is generally no longer regarded as appropriate in contemporary medicine[34]; relatives may request withholding of information about impending death if nothing can be done to avert it – which can lead to tension in the *patient-doctor-family* relationship. In modern practice, disclosing bad news is central to the role of a medical professional[35]–[37] but may be considered one of the more challenging and stressful responsibilities[35]. So, if a specific communication has potential negative consequences for the subject (e.g., stress induced in the healthcare professional) and the object (e.g. distress and reduced life choices for the patient, uncertainty) can it be justified?

Bad news encapsulates a plethora of scenarios and can range in significance from a delayed appointment to a terminal diagnosis. For the remainder of this manuscript, we use the term 'bad news' to refer to a hypothetical situation where impending death is a near certainty for the recipient. Finally, the disclosure of bad news can be analytically sub-divided to include the *decision* to disclose and the *act* of disclosure. We choose to consider the disclosure of *bad news* to be the combined decision and action to communicate the knowledge of bad news to an individual.

**The Felicific Calculus**

Utilitarianism is one of the main branches of *consequentialism* within normative ethics (the philosophical discipline concerned with whether actions are morally right or wrong)[38]. Consequentialists maintain that the morality of an action is determined by its outcome and focus on the *consequences* of a moral act or set of rules[38]. In contrast to other consequentialist theories, <u>*Utilitarianism* values the maximisation of pleasure (or happiness) for the greatest number of people.</u> In healthcare there is a strong *consequence* heuristic, and utilitarianism has become a dominant ethical paradigm: at the public health level, this balances the importance of cost-effectiveness and maximising health benefits for the greatest number of people[15], [16]; and at an individual level explores the risks of a *theoretically* beneficial act, delivering an undesired or even catastrophic outcome. Accordingly, AI has been proposed as a potential moderator of rising healthcare costs along with the having the potential to revolutionise population health[39]. As such, the use of a utilitarian ethical framework to assess the intersection between healthcare and artificial intelligence seems timely.

Put simply, Utilitarian ethicists believe that an action is only 'good' or 'right' if it is productive of the most utility (commonly interpreted as 'happiness' or 'pleasure') compared to its alternatives [14], [40]. Bentham focused his thinking on how the principle of utility could be used practically and considered "pleasure" and "pain" the chief considerations in evaluating happiness (hedonistic utilitarianism). Concisely, he held that the moral content of an act could be seen as a function of the balance between the pleasure and pain that it induces in the subject(s) considered. He devised the felicific (hedonic) calculus to evaluate the balance between the degrees of pleasure and pain that a particular action may cause[14]. An action is assessed through seven different domains: *intensity, duration, certainty, propinquity, fecundity, purity and extent* (Table 1). Thus, the purpose of the felicific calculus is to assist in determining the moral status of an act.

Bentham himself recognised the difficulties of implementing the felicific calculus as a practical tool and considered it most useful for the ethical deliberation of an act. The practical use of the felicific calculus was not intended to be calculative but rather to create a judgement of *good* or *bad* of the pains and pleasures of an action (immediately and subsequently). Nonetheless, attempts have been made to apply *numerical* values to the felicific calculus which has proven to be notoriously difficult[41] and, to the best of our knowledge, a practical working example has never yet been demonstrated. A graphical representation of this concept - *the hedonic scale* (See Figure 1) - has been created for descriptive purposes. The numerical endpoints have been arbitrarily chosen to illustrate positive and negative utility values: '1' representing maximum positive utility (happiness/pleasure), '0' representing no utility value and '-1' representing maximum negative utility (pain). At the end of each section, we declare a tentative score for each domain of the felicific calculus. This scale is mainly illustrative, to aid deliberations on some of the arguments below. The most important question, as per Bentham's formulation of the calculus, is whether the balance between the pleasure and the pain produced by the action falls on the side of the former (>0 on our scale), or the latter (<0 on our scale).

**Figure 1:** Hedonic Scale

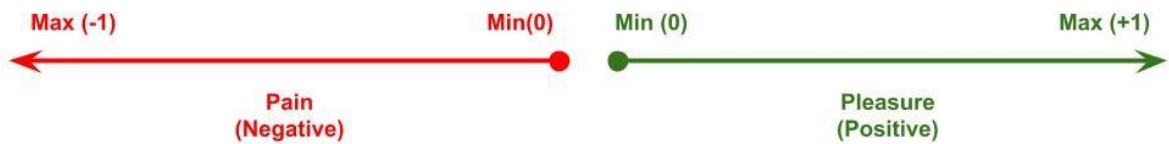

The felicific calculus may not be able to provide a concrete mathematical 'answer' to every moral conundrum, however it provides a useful framework for the deliberation of some philosophical scenarios. Informing an individual that they are dying (*or disclosing bad news*), specifically that their death is impending or imminent, is one such complex scenario. We recognise that a small number of people receiving such information may consider this as a *positive* experience, but for the purposes of this paper, we assume that most people will consider the receipt of this type of bad news an overall negative experience (and for the purposes of this exploration ignore the contrary response). We adopt this stance firstly because, in our opinion, the culturally accepted evaluation of finding out about one's imminent mortality is predominantly negative (shock, dismay, sadness, grief, etc). Moreover, one of the key responsibilities of Western medical practitioners is the preservation of human life[42]. Thus, informing a patient of the impossibility of such preservation may be seen as negative.

In this paper we use *the identification and disclosure of bad news* (impending death) as a test case for a utilitarian analysis in the context of cutting-edge prediction tools. We use this as a 'worked example', or step-by-step solution, for challenging ethical decision-making in modern healthcare.

## 3   Felicific Calculus – *Intensity*

*Definition: The intensity of the pleasure or pain*

> *" Do not go gentle into that good night.*
> *Rage, rage against the dying of the light."*
>                         - Dylan Thomas[43]

Arguably the knowledge of one's own death (mortality salience) is one of the most defining features of the human condition. Despite the inevitability of death, many people find it difficult to discuss[44]–[46], particularly healthcare professionals[47], and even contemplating death leads to avoidance behaviours[48]. Of note however, Walter argues that the discussion of death may not be as taboo as we collectively believe[49]. Learning of one's own impending mortality may be one of the most psychologically traumatising events of a person's life[50]. Dying is not "easy"[51] and it is important to understand that the intensity of emotional pain may vary with time, depending on several factors[52]. Indeed, the behaviour of the agent (doctor), and the manner that news is imparted, can have a significant impact on the trajectory of this pain, but this behaviour only modulates pain and is unlikely to create pleasure[53]–[56].

We believe there is one important positive consequence of breaking bad news: the bestowal of *truth* upon the recipient of the news. After Bentham, George Edward Moore, an *ideal* utilitarian and critic of hedonistic utilitarianism, suggested that the ideals of truth and knowledge are as valuable as pleasure[57]. So, despite truth not being explicitly covered by the felicific calculus, later proponents of utilitarian philosophy, such as Moore, did hold that the maximisation of truth is as important as the maximisation of pleasure[57]. Considering the above, the *intensity* of negativity induced by the knowledge of impending death (and potentially on the agent delivering that information) may be rated *toward* the most negative end of the pain scale (-1). We would note that there are a number of other significant life events (e.g. death of a spouse, divorce) that may induce more psychological suffering than personal illness[58]. In our clinical experience, we have met individuals who believed death would be a welcome relief from intractable chronic suffering such as pain, breathlessness, or loneliness. An interesting study in older people with multiple health conditions demonstrated that a significant number would choose to prioritise independence and pain relief *over* prolongation of life[59]. As such, we have chosen not to assign a fully negative score to *intensity*, but one close to the end of the scale, as many individuals, whilst finding death undesirable, may find alternative burdensome states to be worse.

**Felicific Score: -0.9**

## 4  Felicific Calculus – *Duration*

*Definition: How long the pleasure or pain will last*

Intuitively, estimating the duration of pain should be easy: *the pain lasts until death*. However, there are two major caveats to this premise. Firstly, the grieving process is dynamic, and the intensity of grief may decrease or increase for each individual[60]. Indeed, positive feelings such as relief may even be experienced[61] and members of some religions will see death as a stepping stone to a desirable afterlife or rebirth[62]. Secondly, if attempts were made to minimise the *duration* of pain by delaying or avoiding the disclosure of bad news, the implications could be unpredictable. There is the possibility of 'blissful unawareness' where the patient's life continues unperturbed, no knowledge of impending demise is experienced, and death is sudden, swift and painless. However, more likely is a negative scenario where the person suffers increasing symptomatology combined with an escalating sense of bewilderment and anxiety about their deterioration, with death prefaced only by discomfort, erosion of trust in their physician and eventual realisation. Accurately predicting survival in terminally ill patients is notoriously difficult[60]. Herein lies one, if not the most, significant of the utilitarian dilemmas - it is impossible to know, fully and accurately, the consequence of any action[63].

The stochastic nature of this probability state makes the assignment of a value on the hedonic scale difficult. Rule utilitarianism is one of the main branches of utilitarianism[64], and a rule utilitarian may take the view that one should aim to make the decision that would usually result in the greatest happiness. However, dying itself is a uniquely individual event that can only be experienced by the one dying, leaving aside for the moment secondary consequences on family and other loved ones. In our clinical experience, its trajectory, course, and duration are personal and unpredictable, and thus there is no *usual* dying experience. Furthermore, the durations of the potential pain produced by the anxiety of approaching death and the potential pleasure produced by knowing the truth are both until death. As such we are unable to assign a hedonic value to *duration* in this context and must assume equipoise.

**Felicific Score: 0**

## 5    Felicific Calculus – *Certainty*

*Definition: The probability that the pleasure or pain will occur*

Death is certain[65], but the accurate prediction of time to death (i.e., life expectancy) is difficult, even in those with the most extreme illness[66]–[69]. Under these circumstances there are 3 distinct aspects to *certainty:* the certainty of a prediction (mathematical accuracy, precision, or error), the certainty of the expected outcome (the probability of the outcome occurring) and the certainty that this outcome will lead to pain or pleasure. In this section we do not discuss the certainty of death (we presume death is imminent) but instead focus on the certainty of pleasure or pain emerging from an individual being confronted with the *knowledge* of their impending death, in accordance with Bentham's method.

The receipt of this sort of bad news is generally associated with a "grief" response; the disclosure itself may also cause significant discomfiture for the person imparting the news. Thus, the default position is to assume this action is likely to result in pain. However, it is important to consider that more positive emotions, such as *relief*, may also occur[61], [70]. Indeed, positive and negative emotional reactions may be experienced in succession or concurrently[61]. Utilitarians remain in debate as to whether death is positive (e.g., removal of harm, end suffering, etc) or negative (e.g., prevention of pleasures that would have otherwise been experienced)[71]–[73].

In practice, as with many human-centred theoretical concepts, matters are complicated. For any given individual, the measurement of utility can be difficult and biased. For example, in terminal illness the desire to hasten death is not uncommon[74] but this desire can be confounded by several (potentially dynamic) factors, such as effectiveness of symptom control, mental illness and the dying individual's perceived burden on their family members[75], [76]. Physical pain may be prevalent towards the end of life[77] and the removal of this pain may be seen as 'positive utility'. However, this sensation is entangled with a range of other considerations (e.g., an individual's dignity at the end of life, their current financial affairs, the impact on their family, legal issues e.g., wills, etc)[78]. For any given individual, each of these considerations may be generating positive or negative utility and each factor can be intercorrelated.

On rare occasions a terminal diagnosis may be made in error[79]. One may think that realising a life-limiting illness has been incorrectly diagnosed would result in relief, celebration, and happiness. However, it can have significant negative ramifications including financial[80]–[82] and psychological[83], [84]. This suggests that an individual's response to any life-changing news is deeply personal and based on a plethora of observable and unobservable factors.

Finally, we need to consider how *certainty* is affected by the statistical context of bad news; a terminal diagnosis can be communicated with varying levels of caveats. For example, major trauma can inflict injuries that are not compatible with life e.g., catastrophic brain injuries, so a prediction of imminent death can be accompanied by a high level of certainty. However, many terminal diagnoses (e.g., cancer) may be accompanied by a chance of survival within a given time. For instance, a '1-year survival rate of 50%' can best be understood as: "Historical data suggest that half of people with this diagnosis will still be alive at 1 year". However, for an individual receiving this news, it provides no personalised insight into their life-expectancy. The interpretation of this information will be wholly dependent on the individual – statistical knowledge, personality, outlook, personal experiences, social circumstances, etc. For instance, if a person is diagnosed with pancreatic cancer, their 5-year survival may be as low as 5%[85]. If a patient receives this information any of the following conclusions may be reached and none of them are incorrect:

- "I'm a fighter. I'll get through this and show the doctors I can beat these numbers. I'm going to be in the 5%."
- "I've never been lucky; I'd be surprised if I survived more than one year."
- "Wait, so I have a 95% chance of dying within 5 years? But that could be right now or in 5 years' time! How is this information helpful to me?"

Additionally, every non-deterministic model will have accompanying error rates associated with its predictions. These statistical caveats need to be considered when disclosing bad news, as for example, high false positive rates can lead to unnecessary psychological distress[86], [87]. Furthermore, statistical certainty can impact the certainty of pleasure or pain occurring as inaccurate predictions impact individuals differently[88]. At some point, the precision (or lack of it) of a "prediction" must have an impact on the balance between whether an act of disclosure is *a priori* beneficial or harmful- does it induce *appropriate* or ultimately *inappropriate* pain.

Considering this statistical uncertainty and since neither the clinician nor the patient can have perfect statistical acumen, how can we rate certainty? Note that the 'certainty' we refer to in this section is not the certainty of death, but rather the certainty of the patient's positive/negative experience following the disclosure of bad news. Ultimately, as argued above, we believe the patient's experience will depend on their interpretation of the numbers as well as the numbers themselves. While the downstream utilitarian value is arguably indeterminable, we suggest that the probability of pain being induced by the receipt of bad news is higher than the probability of pleasure. Thus, the felicific score is most likely to be negative, but both potential outcomes are feasible and may occur in the same individual.

**Felicific Score: ≤0 (?)**

## 6   Felicific Calculus – *Propinquity*

*Definition: How soon the pleasure or pain will occur*

As seen above, predicting the timeliness of death is challenging. Healthcare professionals can be asked to predict death on a continuous scale from minutes to years with varying degrees of informative patient-specific knowledge. The communication of medical information, especially if caveated by statistical parameters, is problematic and often leads to confusion [89]–[91]. As such, some individuals, when presented with an estimated survival time, may misunderstand this news and its implications. This added complexity can affect the course of grieving and impact the magnitude, duration and onset of pain or pleasure.

An established grieving paradigm, developed by Kübler-Ross in 1969, outlines stages of grieving that a patient may experience[92]. Kübler-Ross suggests that most people follow a linear journey through 5 stages of grieving: denial, anger, bargaining, depression, and, finally, acceptance. From a utilitarian point of view, this would suggest that in the early stages of grieving, pain and suffering occurs early with the potential for pleasure to occur later. Despite this model of grieving being rejected by modern psychologists[93], intuitively, our deduction that pain is likely to have high propinquity (occurring sooner) and pleasure likely to have a low propinquity (occurring later) after receiving bad news, seems reasonable.

For some, the goal of accurately predicting the point at which a person will die may seem perverse. Could not the benefits of this knowledge only serve to inform administrative tasks such as financial planning, provision of healthcare and resource allocation? Intuitively, the dying person's spiritual, psychological and emotional needs should take priority and it seems feasible that the knowledge of one's death may not serve this aim. We suggest that an individualised approach be adopted, allowing for an individual's preferences to be considered before a time-to-event prediction is made.

In summary, as with our conclusions in section 4, the *propinquity* of a given hedonic sensation can be highly variable. However, we suggest that any suffering is likely to occur soon after the receipt of bad news, whilst pleasure, if it does occur, is much more likely to be delayed.

**Felicific Score: ≤0 (?)**

# 7 Felicific Calculus - *Fecundity and Purity*

*Definitions:*
Fecundity: *How likely the sensation (pleasure or pain) is to lead to more of the same sensation*
Purity: *How likely the sensation (pleasure or pain) is to lead to the opposite sensation*

Once an individual has learned of their impending death, the resulting pain can be persistent and may even increase in severity[74], [94], [95]. However, pleasure can also be experienced during a well-managed end of life event[60], [61] and may, to some extent, be within the person's control[96], [97]. A person's dying experience is individual, with the possibility of experiencing both positive and negative emotions.

The concepts of *fecundity* and *purity* describe how the experience of one pleasurable or painful sensation impacts the likelihood of that same sensation (or the opposite sensation) occurring in the future. We feel that under these circumstances painful sensations are unlikely to lead to pleasure (and vice versa). Overall, it would be difficult to argue that the range of symptoms and emotions that are normally experienced during the dying process are positive or enviable, particularly as these factors tend to deteriorate over time requiring escalating medical intervention[98], [99]. As such, from a *fecundity and purity* perspective the assignment of a number to the hedonic scale must be negative and is likely to be approaching -1.

**Felicific Score: -1**

# 8 Felicific Calculus - *Extent*

*Definition: The number of people affected by the pleasure or pain*

Humans are arguably one of the most successful species on Earth and this is likely, in part, due to our complex social and emotional connectivity and ability to cooperate[100], [101]. Accordingly, events that occur to one individual can have a ripple-like effect on other people, with those closest to that individual experiencing the greatest secondary effects. The dying process has dramatic and unpredictable implications for an individual and few other human events can cause such extensive, rapid and observable downstream effects. The impact on a dying person's family is profound[102]. The loss of a family member is one of the most stressful life events[58] and the resulting grief can increase after death[94], [103]. In fact, in addition to the emotional implications, bereavement can have financial implications[104], physical and psychological implications[105] and even increase the mortality rate in those affected[105]–[107]. Interestingly, a number of studies have shown that, when compared to an *expected* death, the duration and intensity of grieving is worse in those who have *unexpectedly* lost a family member[108], [109]. This is a strong argument in favour of judging the breaking of bad news as consequentially good, as this action may directly reduce suffering. Put another way, while it is hard to argue that the disclosure of the imminent death of a loved one will deliver pleasure, removal of the shock of "sudden death" can mitigate subsequent pain.

The wider impact of death also extends beyond those immediately affected. There is a significant fiscal impact to society[105], [110]. Healthcare utilisation and length of hospital stay increases[105] and these effects may be seen for years[111]. In fact, the people who deliver healthcare may suffer when delivering bad news[112]–[114] and even researchers can be affected by interacting with death[115]. Bereavement has a significant and long-lasting impact on society, but we know that this can be ameliorated, at least in part, by prior understanding and involvement of loved ones [116], [117]. The evidence outlined above suggests that utility to a patient's family and wider society may be maximised if an individual and those around them know that life is drawing to a close. In this context, *extent* should be assigned a highly positive value on the hedonic scale (approaching +1). To clarify, this is not because this action itself implies pleasure, but rather because aiming for less *total* pain demands this action, our argument being a proof by contrapositive: *not* breaking bad news would result in much more pain. A similar example would be the prescription of a medication or vaccination with an unpleasant immediate side effect to prevent a catastrophic disease; occasionally it has been argued that a justification for vaccination is that it delivers a very wide societal benefit. The action induces short-term pain but is directly responsible for avoiding more significant pain in the future.

In truth, assigning *extent* a hedonic score of +1 may be a gross underestimate and arguably should be scaled to account for all the people who have avoided suffering. From a consequentialist point of view the (potential) suffering of the dying individual is outweighed by the amelioration of pain experienced by others. As discussed in Section 2, this action has the secondary benefit of propagating truth across many individuals and institutions. Here we can see the difficulties with using a scoring system such as ours, as intuitively we would opt to assign extent a large positive number, proportional to the number of people impacted. In algorithmic complexity terms this could be considered of the order $O(n)$, where the previous domains would have been $O(1)$. As such, we assert that the positive numerical contribution from *extent's* score should overshadow any negative scores from the first 6 domains due to this qualitative difference, decisively tilting the scales towards the positive side of our evaluation of the rightness of breaking bad news. Thus, we can finally conclude that this act can indeed be justifiable under this normative framework.

It is important to note that it can be inelegant and precarious to justify the suffering of one individual (or group) to maximise the pleasure - or minimise the distress - of another (e.g., Robert Nozick's *Utility Monster* thought experiment[118]) and we maintain that minimising the dying individual's suffering must always be prioritised. Nonetheless, as we have shown herein, the use of a hedonistic framework can be used to demonstrate that breaking bad news is a justifiable act, based on the overwhelming positive impact observed through the lens of *extent*.

**Felicific Score: 1*[O(n)]**

## 9 Discussion

In this example, we have used Bentham's felicific calculus to demonstrate that disclosing the bad news of impending death clearly has negative and painful implications for the target individual, but this action maximises good to *society* by attenuating downstream suffering of others. In addition, the rapidly evolving world of AI, when used for prediction of death (or some similarly adverse or unpleasant event), offers new challenges with potentially unpredictable and harmful effects. The felicific calculus provides one potential framework for deliberating ethical challenges but has numerous practical drawbacks that may limit its use for more complex or uncertain scenarios. In addition, as shown in previous work[16], a utilitarian approach may be useful for healthcare decision-making at a population-level, but inelegant when applied to the individual.

Digital healthcare and medical prediction must always remain patient-centred, and this can only be achieved with trust. As demonstrated during the COVID-19 pandemic, loss of trust when conveying scientific information may lead to counterproductive outcomes or even harm [119]. All individuals associated with the development and communication of healthcare prediction should demonstrate trustworthiness and this is best achieved with 'expertise, honesty and good intentions'[120]. Specific consideration must be given to the way in which we communicate scientific knowledge so as not to mislead our target audience or deprive them of crucial desired additional information[86], [120]. Communicating uncertainty about our knowledge and predictions is vital and can decidedly improve patient trust [88], [120]–[122]. The creation of ground-breaking, complex, and highly accurate AI is futile if the target beneficiaries do not trust its integration in their medical care or comprehend its value and limitations. The reliance on symbol-based media for communication may be error-prone and interpretation-based, especially in the context of AI, where trust very much revolves around such interpretations of language (see the so-called *Interpretation problem*)[123].

Members of the artificial intelligence community are acutely aware of the need to tackle growing concerns over trust and the ethical implications of widespread adoption of this fast-moving technology[124]. However, some have argued that AI itself may be able to resolve its own trust concerns; AI-generated characters have been successfully developed to support education and wellbeing[125]. Steps have even been taken to test AI in highly complex medical communication, with one group deploying virtual agents for end-of-life planning[126]. The conversations with the AI agent were very well-received and the study group of 44 older adults were even comfortable discussing their spiritual preferences with the virtual agent. For some patients, a virtual agent may even be preferred to a real person under some circumstances. In their 2014 paper, Lucas and colleagues demonstrated that some individuals felt more comfortable discussing sensitive issues with a virtual agent, in particular because they felt their responses were not being judged[127]. Whilst much more research needs to be done, early studies suggest that human-AI interactions may be more welcome than might be anticipated.

The pathways for approval of AI for human-use in healthcare is governed by the regulatory category of *Software as a Medical Device[128]*. The most important goal of this process is to minimize use-related hazards and risks so these devices may be used safely and effectively. A key component of this process is *human factors engineering* which includes clarity and precision of communication involving the device. Here, we have shown an approach of how to consider nuanced ethical factors, influencing human interactions with medical AI devices, using the example of predicting and communicating *bad news*.

## 9.1 Limitations, Wider Applicability and Future Work

**Quantitative vs Qualitative**

The assignment of numerical values to a moral or ethical decision rapidly increases in complexity, even if only arbitrarily quantitative once the impact beyond the central individual is considered. With regards to our chosen scoring system, we are limited to an interval of -1 to 1 and thus, cannot capture numerically the expressive power associated with *extent*. This point demonstrates the limitations of numerical analysis for such decisions (compared with qualitative analysis) - a common desire in automating decision-making system[38].

**Plurality through a Coefficient Matrix**

A more comprehensive use of the hedonic scale could include a matrix of hedonic scores for all people affected by an event, with an accompanying coefficient matrix to account for moderator variables such as their 'distance' from an event, numbers of people and individual ages. This would allow a utilitarian value judgement to be assigned to the many people impacted by a dying person and then scaled according to the direct emotional impact to those people the duration of benefit (e.g., younger people may have more suffering ameliorated). However, it could be countered that a coefficient matrix unnecessarily complicates the process as all people will eventually die. Thus, this serves as a natural normalising effect over the course of a person's existence. Regardless, the collection and aggregation of all these values would be extremely complicated and not likely to be available at the point of decision to disclose[38], highlighting the difficulties of practically implementing the felicific calculus.

**Probabilistic Considerations**

We chose to apply the hedonic calculus to a specific set of hypothetical scenarios where two key factors were assumed to be *certain*: the pleasure or harm occurring *and* death occurring. However, this does not reflect reality where the prediction of death is never guaranteed and is accompanied by a level of probability. The same is true for the probability of an outcome (pleasure/harm) occurring. At best, our understanding of future events can only be assigned a probability with an associated distribution of uncertainty. As a result, two layers of (un)certainty then arise: firstly, an *epistemic* probability of a future patient outcome (death in this case, but other "non-death" outcomes clearly exist) occurring and secondly a *consequential* probability of a particular pain or pleasure occurring on learning of this future event. To further complicate matters, these two uncertainties are inherently linked, as the probability and certainty of dying will certainly affect the probability and certainty of harm or pleasure occurring. Importantly, in many circumstances the balance of benefit and harm may be difficult to anticipate as a *probability* clearly acknowledges that a future event may not occur; the distress around receipt of an ultimately incorrect prediction delivers will however be harmful ("pain" only). At a population level, prediction-failure (bad to the individual) must be balanced against predictions success (maybe good overall). This is not a problem unique to AI-based approaches, but many AI-based researchers have an intimate familiarity with probability-based models which acknowledge uncertainty and missing data throughout. Future work should consider how the inherent uncertainty, associated with event prediction, can be accounted for in philosophical analyses.

**Non-binary Prediction Outcomes, Personalisation and the Individual**

Our analysis has considered the implications of bad news where the patient is told they may die, thus the outcome considered is binary - imminent death versus prolonged survival. Other types of bad news (e.g., the diagnosis of a new severe medical condition or a likelihood of a deterioration in health or functional status) would have nonbinary outcome measures that would increase the complexity of a calculus-based ethical analysis. However, it is much harder to envisage, *a priori*, how non-disclosure would be feasible or justifiable under many such circumstances. Individual patient-level preference would need consideration as each person attributes different importance to aspects of their life (e.g., access to their family, independence, mobility). Of note, many standardised quality of life assessments such as the short form 36 (SF-36)[129] and EuroQol-5 Dimension (Eq-5D)[130] instruments do not allow individuals to assign an *importance* to each quality-of-life domain or even identify unspecified "domains" which hold importance to them.

As discussed above in section 4, *Certainty* plays a complicated role in our hedonic calculations. When we consider outcomes other than imminent death, the role of *certainty* is likely to change. For example, disclosing an imperfect prediction of a diagnosis with profound consequences could be catastrophic to an individual. Recent magnetic resonance imaging (MRI) research from the University of Singapore showed that Alzheimer's disease can be diagnosed 4 years before the onset of symptoms, with an accuracy of approximately 80%[131]. For an individual given this information, they would receive a prediction about a disease, with a margin of error, for a disease several years away. In addition, at this point in time, no therapeutic options would be available. Thus, predicting an undesirable future state that is unactionable and uncertain would need careful ethical consideration.

Individuals may choose not to share their bad news with loved ones or may be solitary people without any close personal relationships[132]. Under these circumstances our conclusion, that disclosing bad news is justifiable due to the importance of *extent*, could be challenged. The prevalence of this phenomenon is not well-researched, however our personal experience as clinicians is that this is uncommon, particularly as *dying alone* is a key fear around death for most people[133]. As such, at a population level, this would only apply to a small percentage of individuals.

Given the current acceleration in the development of individual outcome prediction models based on biomarkers, genomic attributes and "big data" derivations, we feel there is a need for further exploration in this domain and future work should consider the real-world application and assessment of the Felicific Calculus or analogous frameworks for breaking bad news. In addition to its practical feasibility, a patient-centred and clinician-centred assessment would need to be undertaken to ensure positive clinical impacts. Note that individualised health (including prediction) is a major focus of health policy and AI development internationally[8], [134].

**Alternative Moral Paradigms**

The felicific calculus itself needs interrogation to assess its validity as a framework to judge clinical ethical queries. Other philosophical and ethical systems (e.g., deontology, virtue ethics) should be used to appraise the ethical justification for breaking uncertain bad news. Western medical ethics have conventionally been based on Principlism (autonomy, non-maleficence, beneficence, justice)[135]. However, this approach is not always well received, especially in cultures where the *family* is the central unit of identity rather than the 'individual'[136], [137]. Instead, *Narrative Ethics* is increasingly being discussed as a more nuanced approach to breaking bad news[138], [139]. Its central tenet being that every ethical dilemma is unique and thus a set of universal ethical rules or principles cannot be applicable to every situation[140]. As we move to more globalised medicine, particularly as technology and information can be rapidly shared internationally, our approach to bioethics may need to change. Non-western countries and cultures can have different foundational moral paradigms, from which locally applicable approaches to bioethical questions arise[141]. Indeed, this subject can be highly complex as ethical paradigms may arise from completely different causal interpretation of disease, healthcare priorities, and cultural preferences[141]–[143]. However, despite different foundations, there can be significant overlap in moral conclusions between western and non-western philosophies[143].

# 10 Conclusion

We have used the felicific calculus as a framework to discuss the utilitarian implications of telling someone they are dying. Communicating bad news will always be challenging and the decision to disclose this information should always be made after careful consideration of each individual and the potential impact on others. The first six elements of the felicific calculus focus on a detailed description and analysis of the pain or pleasure of the target *individual*. As demonstrated above, for each of these *hedonic* elements, our arguments along with knowledge gained from the existing literature indicate that breaking bad news either induces suffering (intensity, fecundity, purity) or an unpredictable outcome (certainty, duration, propinquity). If we only considered the individual, using these six elements in isolation, the action of telling someone they are dying may not be justifiable under a hedonistic framework. However, once the seventh and final element of the felicific calculus, *extent*, is accounted for, we could reach a different conclusion:

Breaking bad news can be viewed as a *good* act when other people and wider society are considered, in addition to the individual, due to the powerful impact of *extent*.

We hope that our analysis will both inform future developments in ethical AI and serve as a proof-of-concept for solving other ethical conundrums in healthcare.

24